\documentclass[]{spie}  %>>> use for US letter paper
\usepackage[]{graphicx}
\usepackage{amsmath}
\title{Characterization of bent crystals for Laue lenses} 

\author{V. Liccardo\supit{a,b}, E. Virgilli\supit{a}, F. Frontera\supit{a},
V. Valsan\supit{a,b}
\skiplinehalf
\supit{a}\textit{Physics Department, University of Ferrara, Via Saragat 1, 44122 Ferrara, Italy;}
\supit{b}\textit{Universit\'e de Nice Sophia Antipolis, Nice, Cedex 2, Grand Chateau Parc Valrose, France; }\\
}

\authorinfo{Further author information: (Send correspondence to:)\\
E-mail: virgilli@fe.infn.it, frontera@fe.infn.it, liccardo@fe.infn.it,
valsan@fe.infn.it}
\begin{document}  
  \maketitle 

%%%%%%%%%%%%%%%%%%%%%%%%%%%%%%%%%%%%%%%%%%%%%%%%%%%%%%%%%%%%% 
\begin{abstract}
In the context of the LAUE project devoted to build a long focal-length focusing optics for 
soft $\gamma$--ray astronomy (80 -- 600 keV), we present the results of reflectivity measurements of 
bent  crystals in different configurations, obtained by bending perfect or mosaic flat crystals. 
We also compare these results with those obtained using flat crystals. The measurements were performed 
using the K$\alpha$ line of the Tungsten anode of the X--ray tube used in the LARIX facility of the University of Ferrara. 

These results are finalized to select the best materials and to optimize the thickness of the crystal tiles
that will be used for building a Laue lens petal which is a part of an entire Laue lens, with 20 m focal length and 
100--300 keV passband. The final goal of the LAUE project is 
overcome, by at least 2 orders of magnitude, the sensitivity limits of the current generation of $\gamma$--ray telescopes, and to improve the current $\gamma$--ray imaging capability.  
\end{abstract}

%>>>> Include a list of keywords after the abstract 


%%%%%%%%%%%%%%%%%%%%%%%%%%%%%%%%%%%%%%%%%%%%%%%%%%%%%%%%%%%%%
\section{INTRODUCTION}
 \label{sec:intro}  %\label{} allows reference to this section

Experimental hard X--/soft $\gamma$--ray (10-1000 keV) astronomy is moving from direct sky viewing 
telescopes to focusing telescopes. With the forthcoming focusing telescopes in this energy range, a big 
improvement in sensitivity is expected: a factor of 100-1000 with respect to the best non-focusing 
instruments of the current generation, either using coded masks or not. A significant increase in angular 
resolution will be also achievable from the $\sim$~10 arcmin of the mask telescopes to less than 1 arcmin. 
The first generation of soft $\gamma$--ray ($>$~100 keV) focusing telescopes will make use of the Bragg diffraction 
technique from crystals in a transmission configuration (Laue lenses).

The astrophysical issues that are expected to be solved with the advent of these telescopes 
are many and of fundamental importance (see, e.g., Ref.~\citenum{Frontera10}).
We have already developed two Laue lens prototypes, in the framework of the Hard X-ray 
TELescope (HAXTEL) project, devoted to develop and test a technology for building a broad  
passband Laue lens for small focal 
length ($<$10 m){Frontera08,Virgilli11}.
After the successful results of HAXTEL, a new LAUE project is ongoing, devoted to build broad band (80--600 keV) focusing lenses with longer focal lengths.
The final goal of the LAUE project is to build a lens with flux high sensitivity. This implies that the focal spot has to be as small as possible, not only to improve the angular resolution but mainly to reduce the  background under the spot. To achieve this goal, we moved from flat mosaic crystals employed in the HAXTEL project, 
to bent crystals. The are provided by the "Laboratorio Sensori e Semiconduttori (LSS -- Ferrara)" and 
by "Istituto dei Materiali per l' Elettronica ed il Magnetismo (IMEM -- Parma)", both institutions involved in the LAUE project.

In this paper, after having shortly described the LAUE project, a description of the bent crystals tested and their properties is given with a comparison of the corresponding properties of flat perfect and mosaic crystals.
Finally, the experimental results concerning bent Si and GaAs crystal samples are reported.

\section{THE LAUE PROJECT}
\label{sec:laueproject}

The Laue project, supported by the Italian Space Agency (ASI), is devoted to development of an advanced 
technology for building a Laue lens  with a broad energy passband (80--600 keV) and long focal length 
(up to 20 m and beyond) for space Astrophysics (for a detailed overview of the entire
project see Ref.~\citenum{Virgilli11,Frontera12}. From a technical point of view, the massive production of 
crystals with the   
demanded characteristics, and the development of a technology for a fast positioning with an accuracy better 
than 10 arcsec, represent a tough challenge. Only a very tight collaboration between scientific institutions and
 and industries has made possible to employ the most advanced technologies in crystals manufacturing and the ongoing installation of a new facility, in which the crystals will be tested and 
assembled, with the required accuracy,  on a lens petal. The adopted technology for assembling the crystals in the lens is to fix the crystal tiles on its place on the lens frame after properly orienting them under the control of a $\gamma$--ray beam. The correct fixing of the crystal tiles to the lens frame is ensured by gluing 
them to the frame kept in the same position during the assembling phase \cite{Virgilli11b}.
Given this assempling strategy, both the X-ray source and the collimator have to be moved together in order that the beam axis is parallel to the lens axis and all crystal tiles can be properly fixed to lens frame. Such a procedure ensures that each crystal focuses the incident photons in the lens focus, within the equipment uncertanties.

As a result of the LAUE project, a lens petal 
with 20~m focal length and 100--300 keV passband will be built.

\section{CRYSTAL TEST APPARATUS}
\label{sec:larix}

\begin{figure}[!h]
\begin{center}
\includegraphics[scale=0.3]{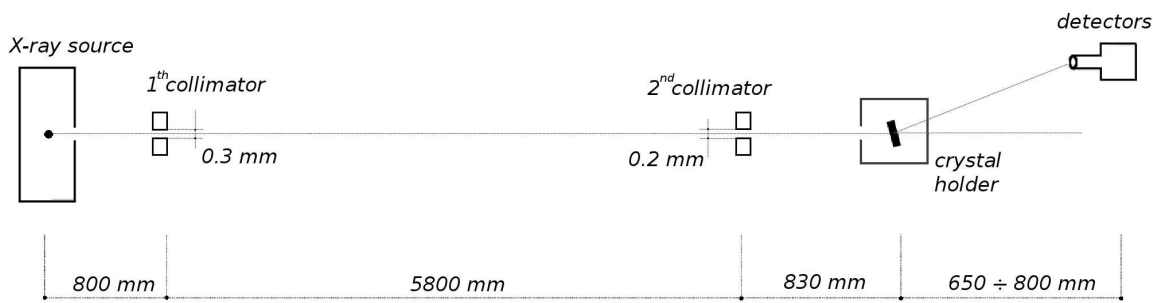}
\caption{\footnotesize Sketch of the top view of the apparatus used for the crystal testing.}
\label{fig:facility}
\end{center} 
\end{figure}

The test of samples of candidate crystals for the LAUE project
has been performed with the small LARIX facility shown 
in Fig.~\ref{fig:facility}. By means of a motorized crystal holder, it is possible 
to translate the crystal samples along 2 directions perpendicular to the beam and rotate them around 
three orthogonal axes. The X-ray beam coming from the source (an X-ray tube with a maximum voltage of 150 kV) travels across 
two collimators with adjustable size, the former at 80~cm from the X-ray source, the latter at a distance of 83~cm
from the crystal to be 
tested. The distance between the collimators is 580~cm.
The direct and reflected beam is analyzed by means of two detectors, an X-ray imager 
with spatial resolution of 300 $\mu$m, and a cooled HPGe spectrometer with a 800 eV spectral resolution 
at 100 keV, both positioned on a detector holder that can moved back and forth along the beam axis \cite{Frontera08}. 

Figure 1

\section{MEASUREMENT SET-UP}
\label{sec:experimental}

\begin{figure}[!h]
\begin{center}
\includegraphics[scale=0.2]{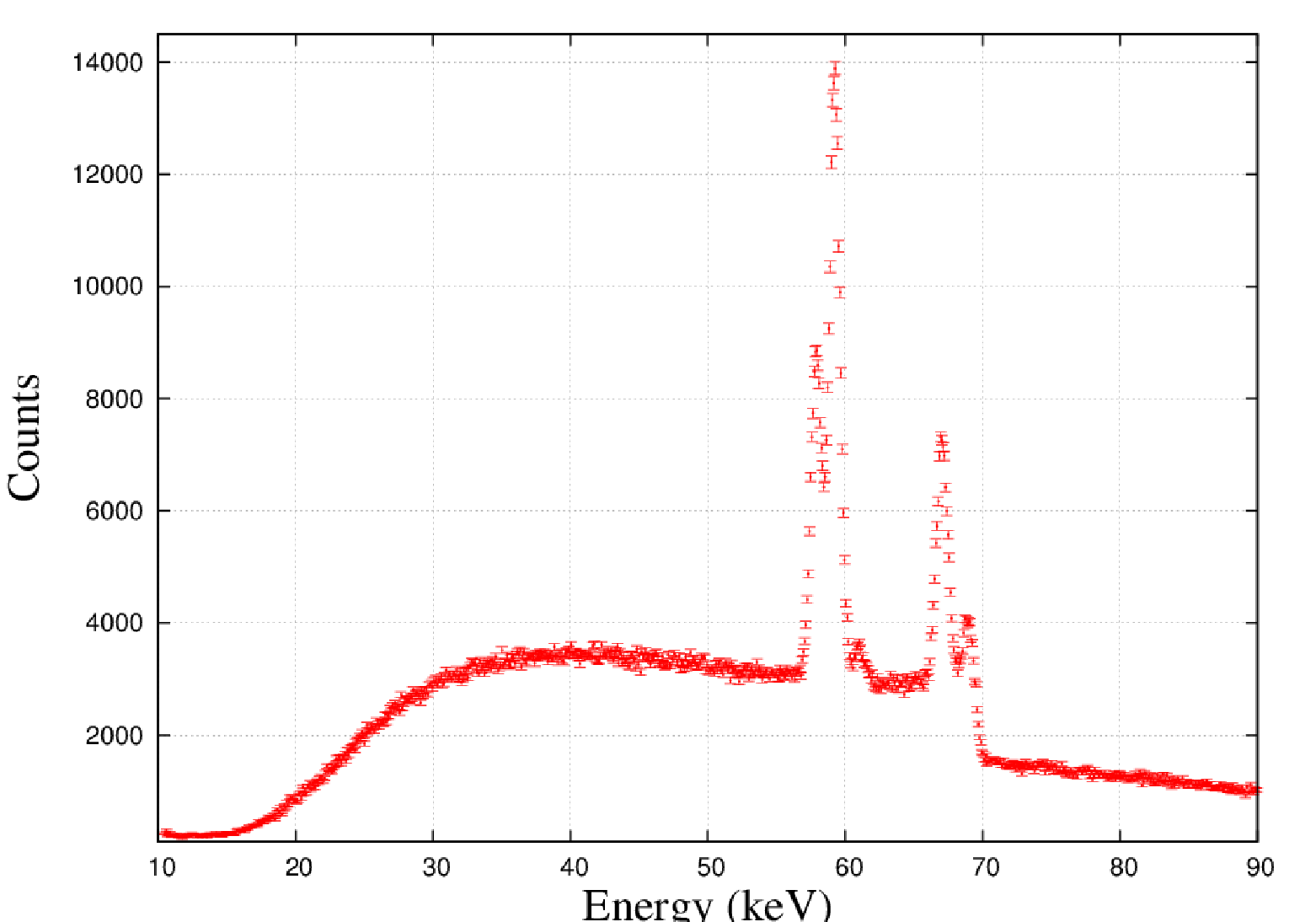}
\includegraphics[scale=0.2]{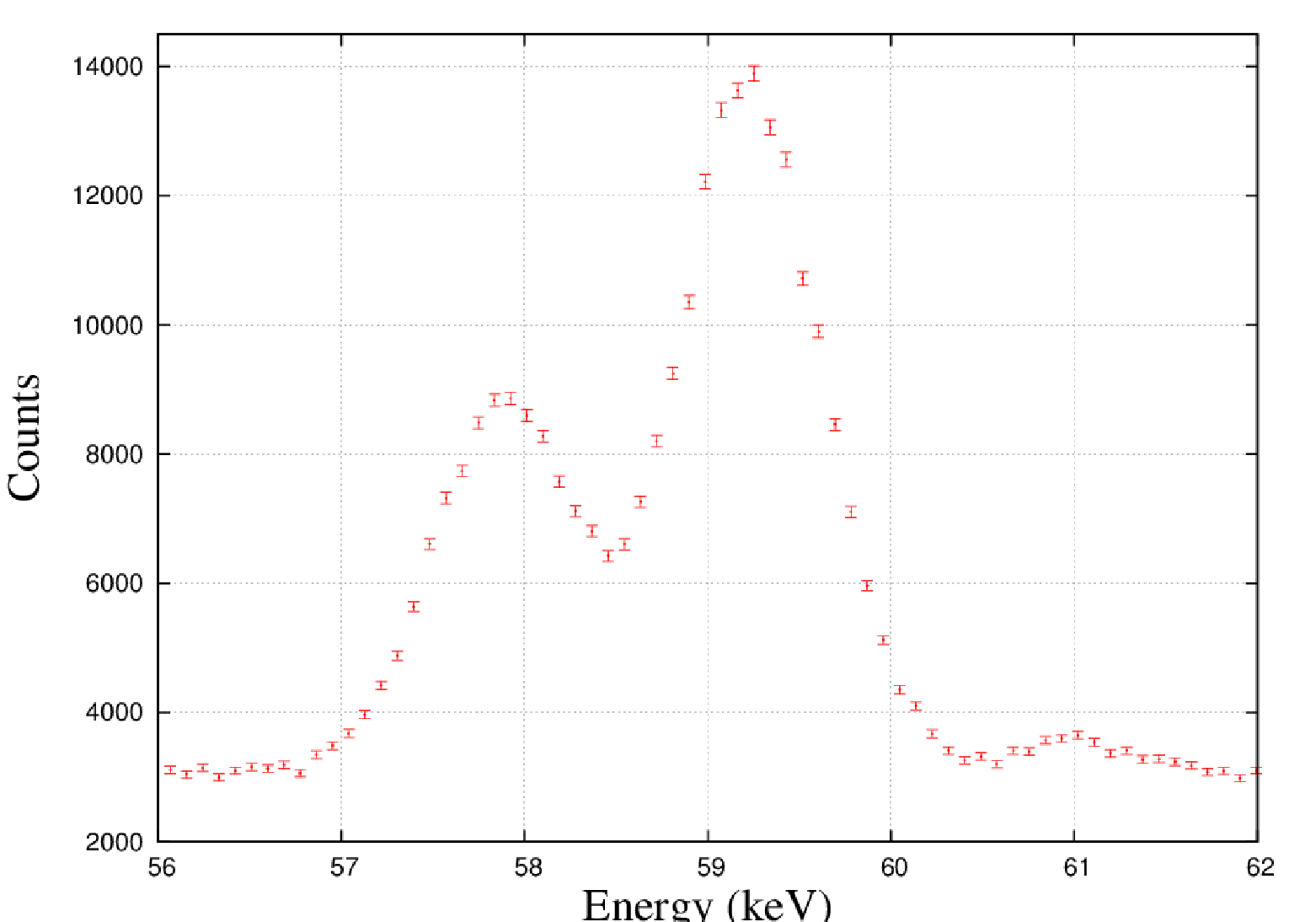}
\caption{\footnotesize $Left:$ X-rays tube polychromatic spectrum. $Right:$ K$\alpha$ line spectrum employed 
for the measurements after removing the Bremsstrahlung background.}
\label{fig:line}
\end{center}
\end{figure}

The angular distribution of the diffracting planes and the main properties of each sample  
are obtained by measuring the FWHM of the Rocking Curve (RC) and the reflectivity value.
The measurements have been performed using a monochromatic beam at 59.2 keV obtained from the fluorescence K$\alpha$ line
of the Tungsten anode of the  X-ray tube 
(Fig.~\ref{fig:line}).

%
% Figure 2
%

The limited length of the experimental set-up (see Fig.~\ref{fig:facility}) and the X-ray source size (0.6~mm radius) give rise to a divergent beam impinging on the crystal sample. An accurate estimate of this divergence is thus  crucial, to separate this effect from the response 
function of each tested crystal. 

A divergence measurement has been carried 
out, using a flat Silicon crystal (111) in Bragg configuration, by measuring the rocking curve of the crystal 
for different collimator sizes and comparing it with the theoretical value, geometrically 
estimated.

\begin{figure}[!h]
\begin{center}
\includegraphics[scale=0.2]{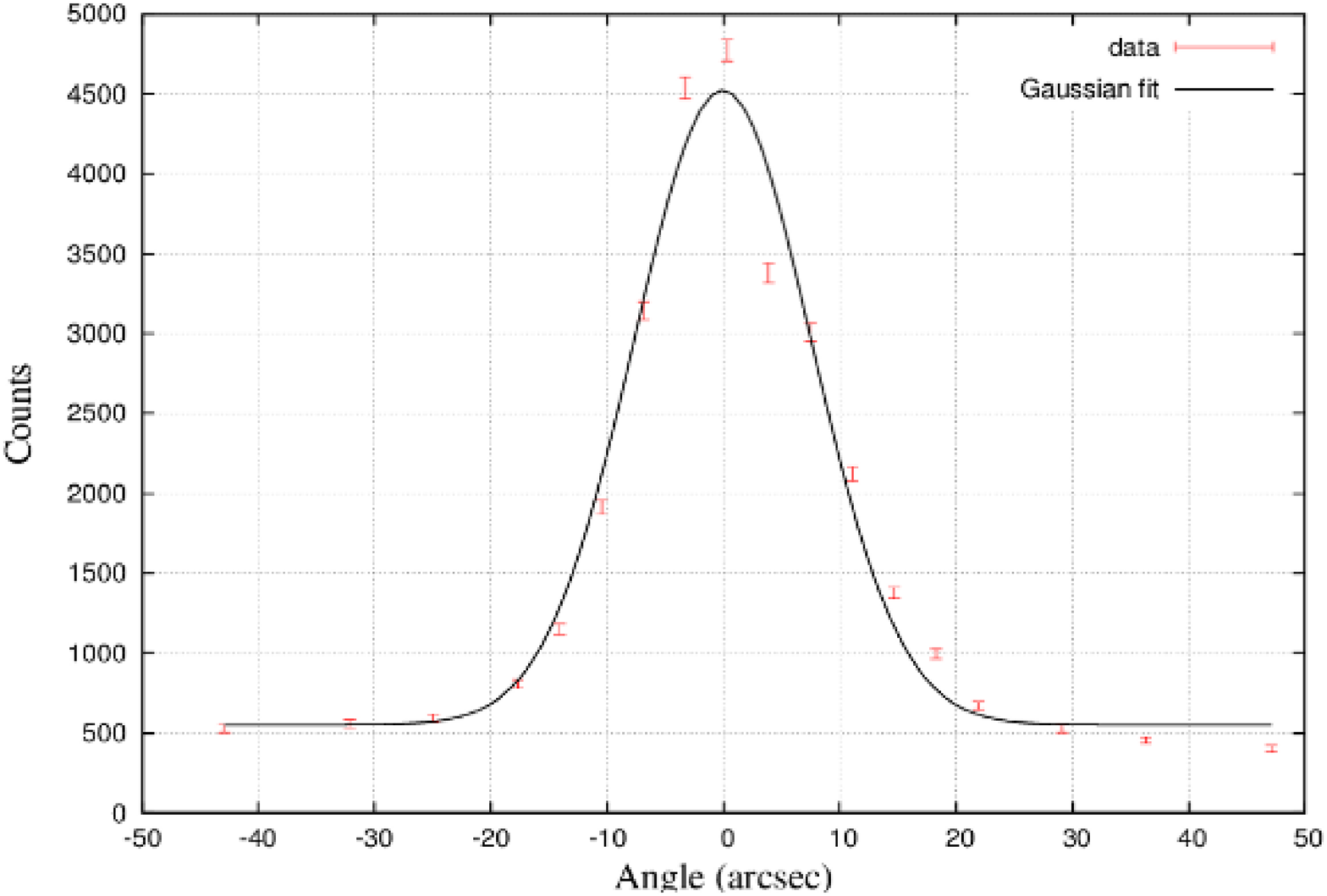}
\includegraphics[scale=0.25]{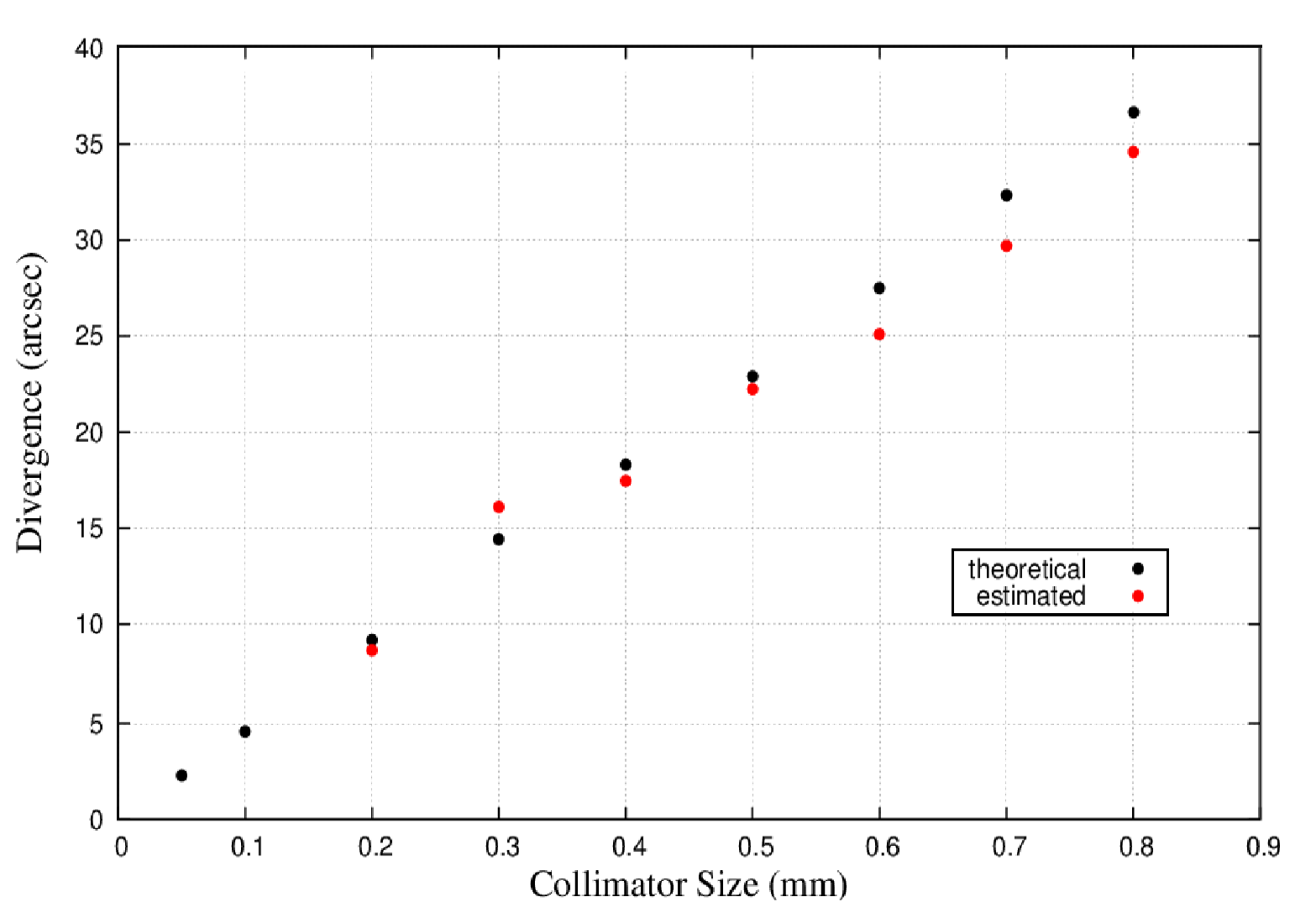}
\caption{\footnotesize $Left:$ Beam divergence with a collimator aperture of 0.3 $\times$ 0.2 mm. 
$Right:$ Geometric value of the divergence as a function of the collimators aperture.}
\label{fig:divergence}
\end{center}
\end{figure}

A compromise between brightness of the beam that is impinging on the crystal and divergence level has been reached, by setting the horizontal aperture of the collimator 1 to 0.3 mm and that of the collimator 2 to 0.2 mm,
keeping the vertical aperture of both collimators to 2 mm.
With this configuration the resulting divergence value (Full Width at Half Maximum, $FWHM$)  was estimated to be $FWHM_{\rm div} = 11 \pm 2$~arcsec, in agreement with 
the geometrical estimate (see Fig. \ref{fig:divergence}).

%
% Figure 3 
%

In order to estimate the sample reflectivity of perfect crystals at 59.2 keV, we compare the incident beam intensity (naturally integrated over all divergence angles) with the crystal Rocking Curve, obtained by rotating the sample in Laue configuration. 
For the mosaic samples, the effective $FWHM$ of the mosaic spread (crystal mosaicity $\eta$, see Ref.~\citenum{Frontera10}) is instead obtained from the Gaussian spread $FWHM_{\rm meas}$ of the measured diffraction peak through the following 
equation \citenum{Ferrari11}:
\begin{equation}
FWHM  = \sqrt {FWHM_{\rm meas}^2 - FWHM_{div}^2} 
\label{eq:sigma}
\end{equation}

\section{CRYSTAL SAMPLES and TESTED GEOMETRIES}

The crystal samples (see Table~\ref {tab:table}) have been tested by setting the X--ray tube voltage to 120 kV 
with a current of about 1.2 mA. For each sample, both the intensity of the trasmitted beam through the crystal and the Rocking Curve  
are measured. The direct beam is also periodically monitored for possible intensity variations with time.

Two configurations of the crystals have been investigated: $geomtery~1$ and $geometry~2$. In $geometry~1$ the beam is impinging on the lateral side of the crystal tile, i.e., the susface in which one of the two sizes is the crystal thickness. Instead, in $geometry~2$, the X--ray beam is impinging on the main face of the crystal tile.
The $geometry~2$ is used for testing the flat and bent mosaic crystals, and  the bent perfect crystals in Laue configuration, where a secondary internal curvature of the lattice planes (quasi-mosaic)
arises due to the induced external bending.

Two crystal holders are available. The first is a clamp support, which is employed for testing squared or 
rectangular crystals tiles 10 $\div$ 20 mm long each side and 0.5 $\div$ 3 mm thick.
This crystal holder is used for $geometry~2$ configuration. 
The other holder is suitable for testing crystals in $geometry~1$. 
The holder is made of a steal base, on which it is positioned  an aluminum wall (see right panel of Fig.~\ref{fig:groove})that can be moved back and forth along 
the base depending on the crystal dimensions, in order to rotate the analyzed crystal around the axis of the holder base and to get it within the beam axis.

\begin{table}
\caption{\footnotesize Tested samples} 
\label{tab:table}
\centering
\medskip     
\begin{tabular}{c|c|c|c} 
Number &  Material &  Dimensions & Geometry\\
  of tiles & & $mm$ x $mm$ x $mm$ &\\
  
\hline
 1 & Silicon & 15 x 15 x 0.75 & Flat perfect\\
 1 & Silicon & 15 x 15 x 0.75 & Perfect, bent 60 m curvature radius\\
 1 & Silicon & 15 x 15 x 0.75 & Perfect, bent 27 m curvature radius\\
 1 & Silicon & 10 x 15 x 10.5 & Stack of 14 tiles\\
 5 & Gallium Arsenide & 15 x 15 x 2 & Flat mosaic\\
 1 & Gallium Arsenide & Rounded shape x 2 & Bent mosaic\\
\hline
 1 & Silicon & 25 x 25 x 1 & Flat\\
 1 & Silicon & 25 x 25 x 1 & Bent via indentation\\
 1 & Silicon & 45 x 10 x 3 & Stack of 3 elements, bent via indentations \\
\end{tabular}
\end{table}

\section{RESULTS}

\subsection{Samples provided by IMEM}

Bent crystals are obtained by lapping one of the main crystal cross sections with sandpaper that 
introduces defects in a superficial layer undergoing a highly compressive strain. The samples 
provided by IMEM are shown in the upper part of Table~\ref {tab:table}.

\subsubsection{Flat Silicon, Bent Silicon 60 m and 27 m curvature radius in geometry 2}

\begin{figure}[!h]
\begin{center}
\includegraphics[scale=0.25]{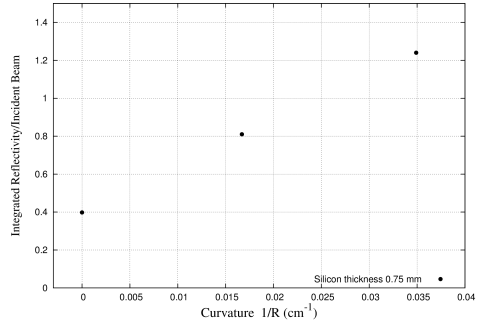}
\caption{\footnotesize Plot of the Integrated Reflectivity normalized to the incident beam with respect 
to the curvature of the crystal for different curvature: R = $\infty$, R = 60 m, R = 27 m}
\label{fig:curvature_parma}
\end{center} 
\end{figure}

Both the flat Si sample and the two bent Si samples have a dimension 15 x 15 x 0.75 mm$^3$ with diffracting planes in $geometry~2$ being  the (100). They all come from 
the same ingot, therefore the measurements provide a good comparison of their properties.
As shown in Fig. \ref{fig:curvature_parma}, the  reflectivity at 59.2 keV  increases 
linearly as a function of the crystal curvature, demonstrating that the lapping technique 
can improve the diffraction efficiency of a flat perfect crystal. For a 60 m and 27 m curvature 
radius the reflection efficiency increase by a factor 2 and 3, respectively, with respcet to that of the flat crystal.

%
% Figure 4
%

\subsubsection{GaAs flat mosaic crystal in transmission configuration} 

\begin{figure}[!h]
\begin{center}
\includegraphics[scale=0.12]{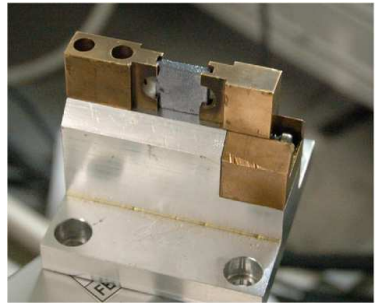}
\caption{\footnotesize GaAs flat mosaic sample mounted on the support.}
\label{fig:gaas}
\end{center}
\end{figure}

Two samples were tested, using the clamp crystal support (Fig. \ref{fig:gaas}). The tested samples 
are known to have a mosaicity of about 25 arcsec.  Their cross-section is square with dimensions 
of 15 x 15 mm$^2$ and thickness of 2 mm.  The measurements are found to be consistent with the expectations \cite{Ferrari11}, with 
a value of the reflectivity at 59.2 keV of about 35\% and a mosaicity of about 25 arcsec.

%
% Figure 5
%

\subsubsection{Bent mosaic crystal of GaAs(111) in transmission configuration} 

\begin{figure}[!h]
\begin{center}
\includegraphics[scale=0.2]{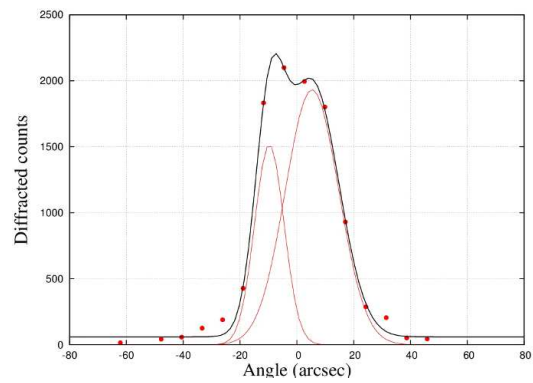}
\caption{\footnotesize RC of the GaAs sample (40 m radius) bent crystal.}
\label{f:gaas}
\end{center}
\end{figure}

The sample has a rounded shape with a radius of 35 mm while the bending radius is
40 m. We mounted the tile on a 
support and analyzed it in the central region in Laue configuration, being (111) the diffracting planes. 
The measured efficiency (see Fig.~\ref{f:gaas} is about 40\%, slightly greater than the flat sample, consistent 
with a small increase of the angular spread with respect to the flat GaAs crystal (25--30 arcsec).
The curvature does not affects the internal structure and the local mosaic spread does not change, 
while it influence the global behaviour of the sample that allows the focusing effect. 

%
% Figure 6
%

\subsubsection{Stack of bent Silicon (111) crystals in geometry 1} 
\label{sec:stack}

\begin{figure}[!h]
\begin{center}
\includegraphics[scale=0.1]{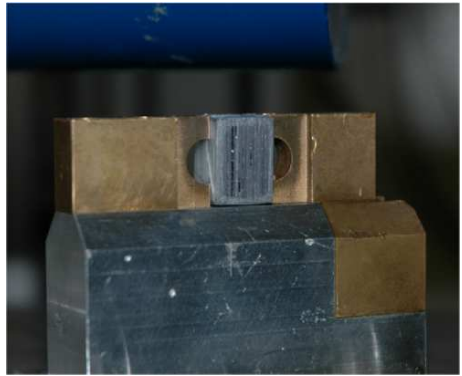}
\caption{\footnotesize The Silicon stack on the crystal support during the test.}
\label{si_stack}
\end{center}
\end{figure}

We have also tested a stack of 14 Si(111) crystals (Fig.\ref{si_stack}) in $geometry 1$ with a bending radius of 
55 m. Each crystal has a size of 10 $\times$ 15 mm$^2$ and a thickness of 0.75 mm, with the stack showing a surface, on which the beam is impinging, of $10\times10.5$~mm$^2$.

%
% Figure 7
%

The sample has been analyzed in two different regions.  Although the stack was assembled 
manually and the tiles were not glued among them, it showed similar diffraction properties, with a peak efficiency at 59.2 keV of about 30\%.

Although  stacks made of bent silicon have been discarded due to  
technical difficulties in producing a large quantity of stacks with the desired alignment, 
the opportunity of testing the crystals in this configuration is equally important for possible future applications.

\subsection{Crystals provided by LSS}
\label{sec:lss}

The LSS laboratory at the University of Ferrara provided us bent Silicon crystals with lattice 
planes (111) perpendicular to the main crystal surface. The bending is achieved by indentation of 
a mesh of grooves of one of the surfaces 
of the crystal \cite{Camattari11}. 
The obtained curvature is permanent
and can be finely tuned by changing the gooving parameters, such as blade features and grooving speed, geometry 
and size of the grooves. It can be demonstrated that for some of the lattice planes used as diffracting planes in Laue configuration, when a primary 
curvature is provided to the crystal, a secondary (quasi-mosaic) curvature of the diffracting planes 
occurs (Fig.~\ref{f:quasi}). This secondary curvature can be properly exploited for increasing the diffraction efficiency.
The final bending is verified by means of a profilometer with an uncertainty 
of 5\% (2 meters in our case, being the lens curvature radius 40 meters).

%
% Figure 8
%
\begin{figure}[!h]
\begin{center}
\includegraphics[scale=0.06]{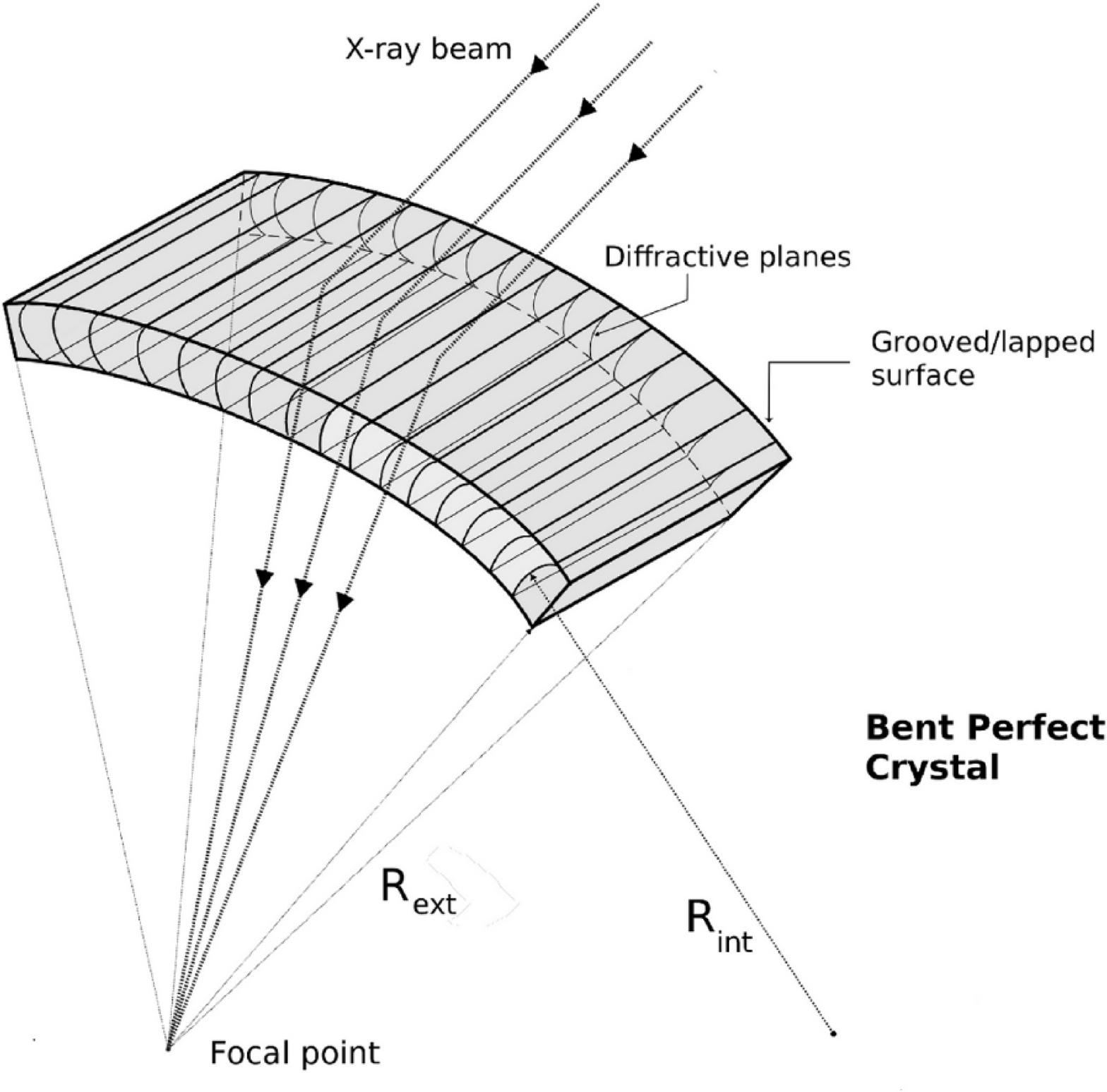}
\caption{\footnotesize Sketch of a bent perfect crystal in quasi-mosaic geometry.}
\label{quasi}
\end{center}
\end{figure}

\subsubsection{Silicon bent crystals in geometry 1}

The sample has a volume of $25 \times 25 \times 1$~mm$^3$, with 1 mm being its thickness. It was analyzed in geometry 1,  with the 
beam penetrating the crystal through the lateral surface 
$25\times 1$~mm$^2$, perpendicular to the grooved plane.

Table~\ref{tab:table2} shows the FWHM of the RCs obtained rotating the crystal around the axis
of the main surface of the crystal, taking into account the divergence effect of the beam. A relation seems to be
present between radiated point position and FWHM, maybe due to the fact that photons impinging on the crystal 
cross-section close to the edges can escape from the lateral sides instead of escaping from the back surface, giving rise to a different angular spread.    

%
% Figure 9
%
\begin{figure}[!h]
\begin{center}
\includegraphics[scale=0.12]{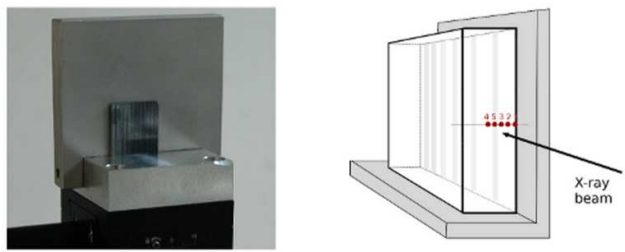}
\caption{\footnotesize $Left:$ Silicon bent crystals analyzed in geometry 1. $Right:$ Pattern of the crystal regions tested.}
\label{fig:groove}
\end{center}
\end{figure}

The results are consistent with previous tests performed by LSS at the ESRF synchrotron radiation 
facility in Grenoble, confirming that LARIX facility is suitable to determine, with good confidence, 
the crystals diffraction properties.

\begin{table}[!h]
\begin{center}
\caption{\footnotesize Rocking curve width as a function of the tested crystal region (Fig. \ref{fig:groove})} 
\label{tab:table2}
\centering
\medskip     
\begin{tabular}{|c|c|c|}
Analyzed & Distance with respect & FWHM  \\   
region                & to the grooved face (mm) & (arcsec) \\   
\hline 
 1 & 0.03 & 25.4 \\
 2 & 0.22 & 43.9 \\
 3 & 0.32 & 51.6 \\
 4 & 0.52 & 52.2 \\
 5 & 0.43 & 67.8 \\
\hline 
\end{tabular}
\end{center}
\end{table} 

\subsubsection{Stack of bent Silicon (220) crystals}

The stack is composed of three Si bent crystals with 110 m bending radius, a front surface of 45 $\times$ 10 mm$^2$ and thickness of 3 mm. 
The corresponding plane curvature is 84.4 arcsec. 
The stack was tested in order to misalignments of single crystals of the stack with each other.

%
% Figure 10
%
\begin{figure}[!h]
\begin{center}
\includegraphics[scale=0.25]{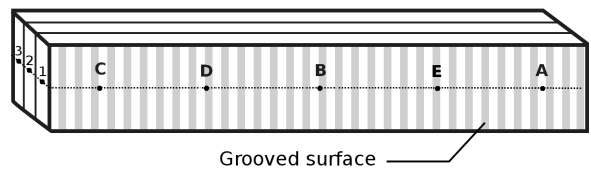}
\caption{\footnotesize  The Silicon stack and the regions tested.}
\label{fe_stack}
\end{center}
\end{figure}

\textbf{Geometry 1 test}\\\\
%\label{sec:geometry1}
The diffractivity of the (111) planes ($\theta_b$$\approx$ $1.92^\circ$) was tested in different points of the 
lateral surface $10\times 3$~mm$^2$ (see Fig.~\ref{fe_stack}).

\begin{figure}[!h]
\begin{center}
\includegraphics[scale=0.2, angle=-90]{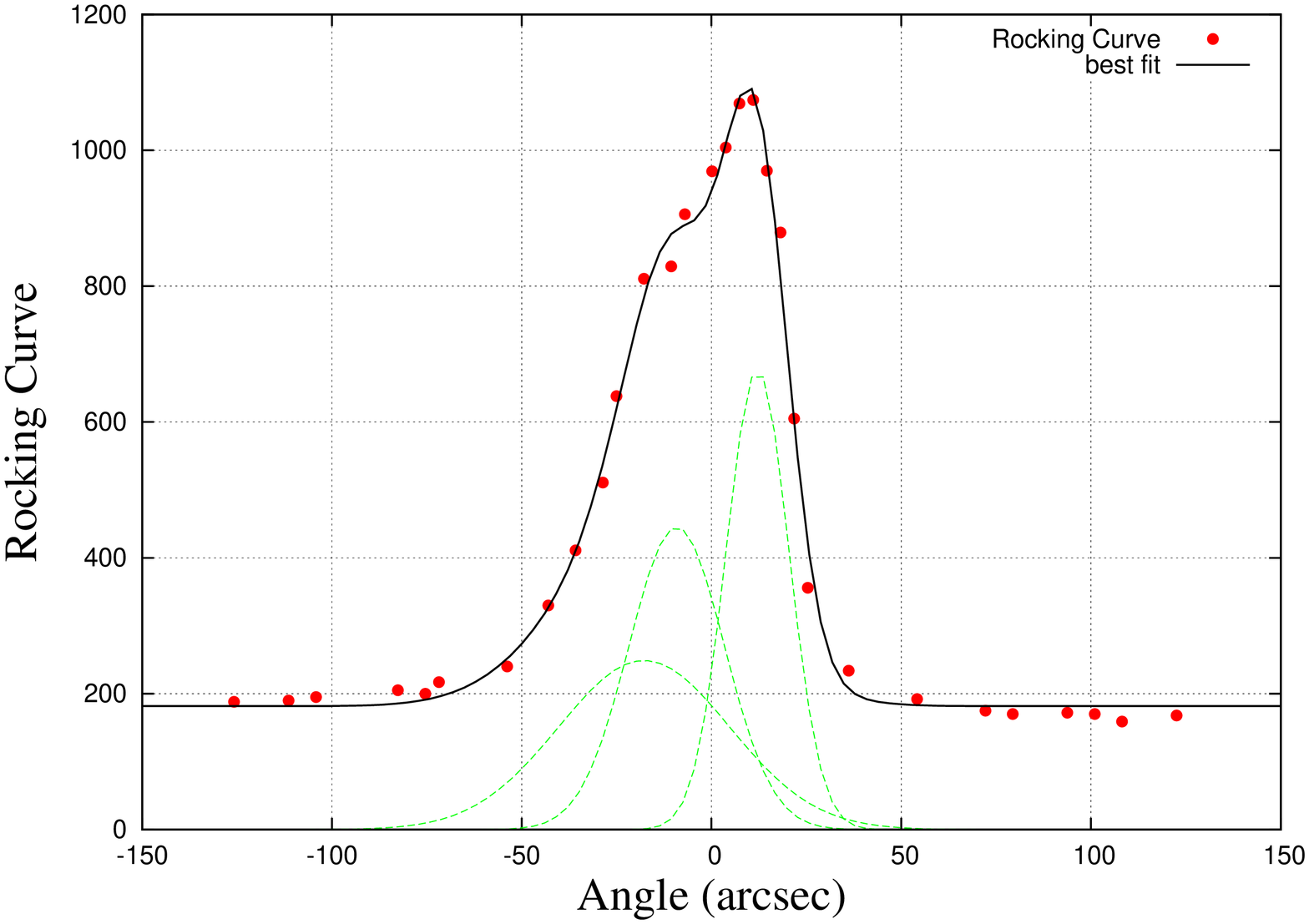}
\includegraphics[scale=0.2, angle=-90]{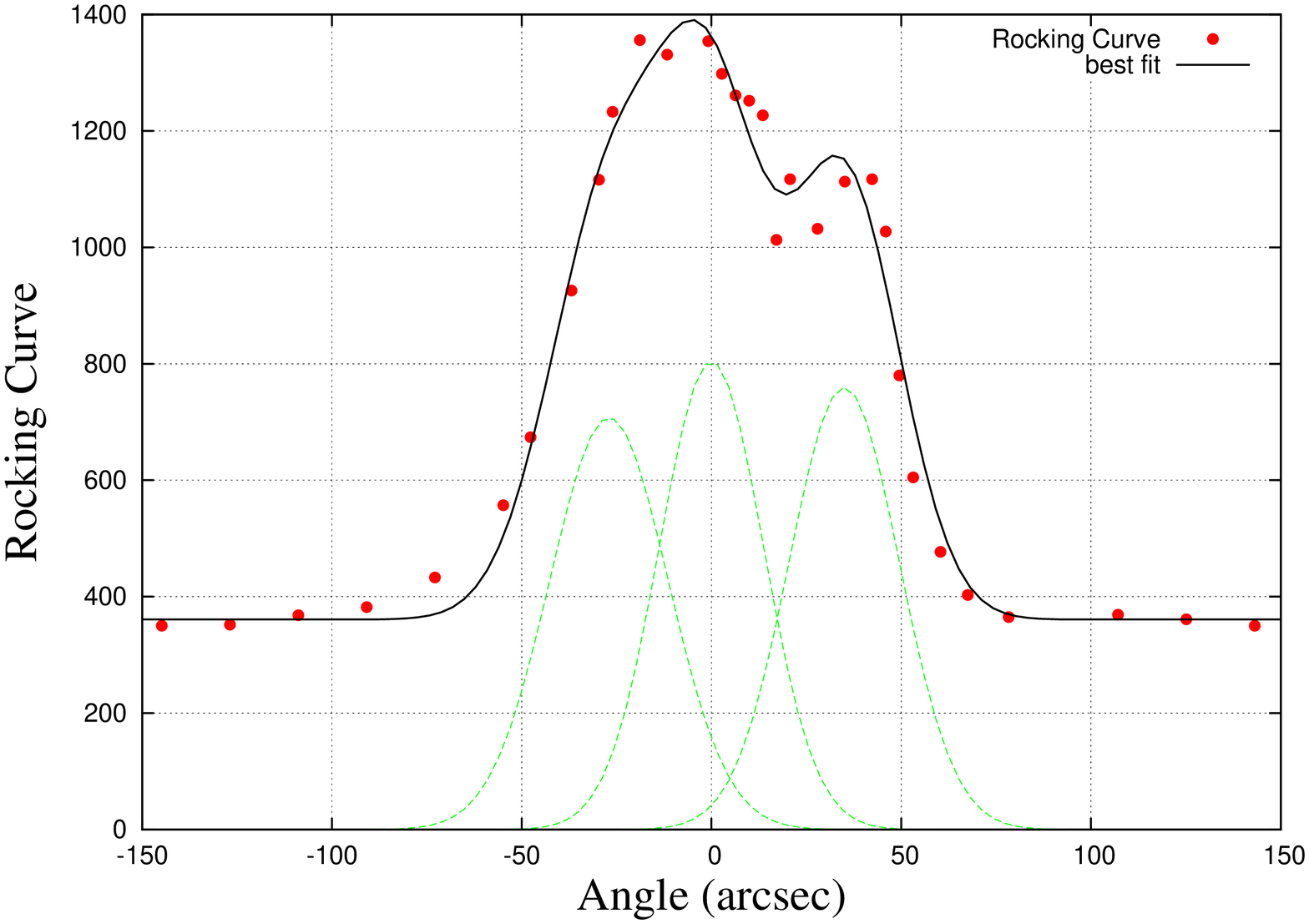}
\includegraphics[scale=0.2, angle=-90]{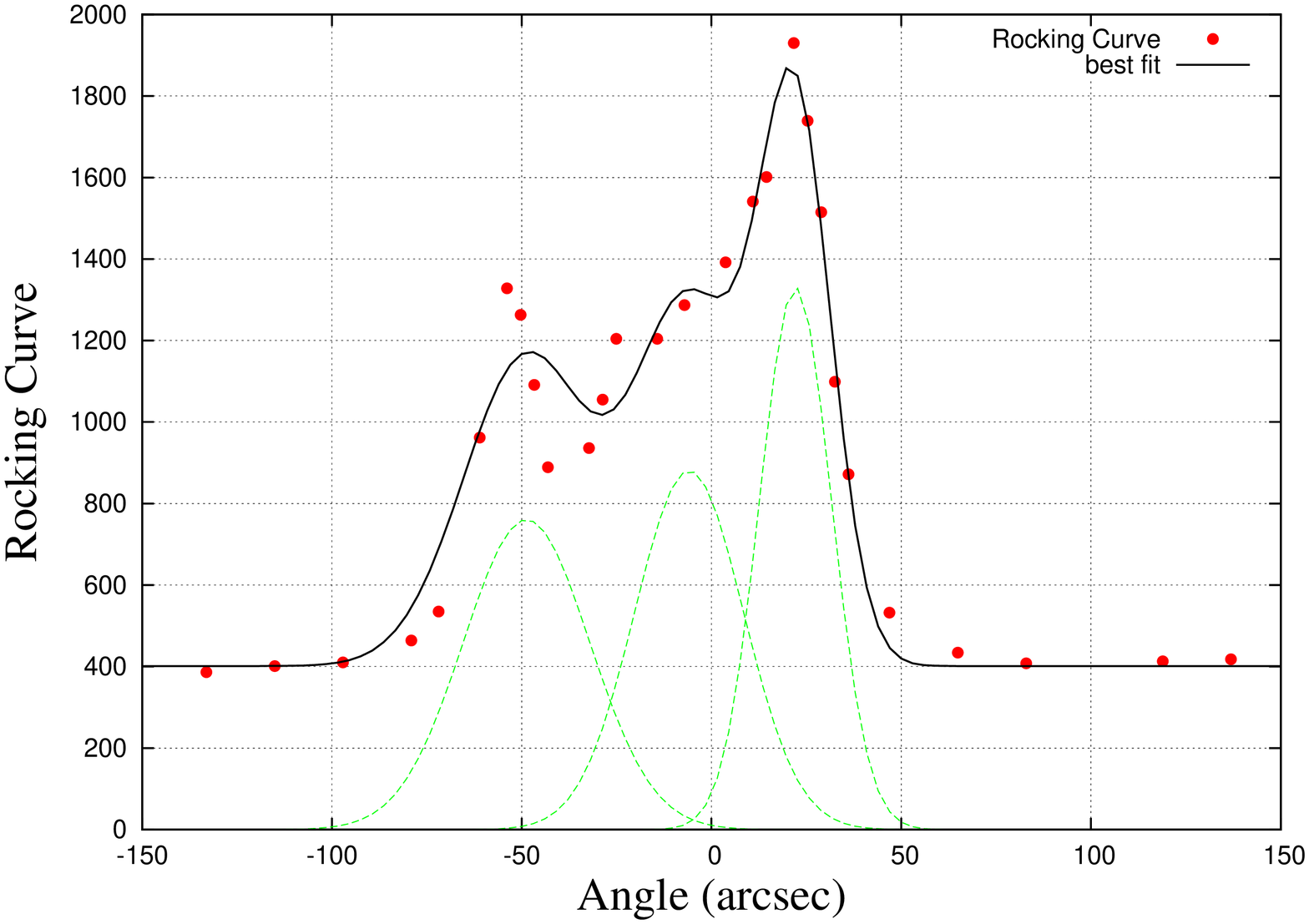}
\newline
\caption{\footnotesize RCs (Geometry 1) of the Silicon stack measured in the regions 1 (left), 2 (center), and 3 (right). The RCs have been fit using 3 Gaussian functions, due to the imperfect alignment of the crystals in the stack and to the depth traveled by the diffracted beam across the layer tested.}
\label{fig:rc}
\end{center}
\end{figure}

The incident X--ray beam enters the stack at different distances from the grooved surface (regions 1, 2, 3, see Fig.~\ref{fe_stack} and Table~\ref{tab:table3}). 
The measured RCs are very broad and are consistent with the sum of three Gaussian (Fig. \ref{fig:rc}). We expected to see a broad peak with FWHM of the order of the crystal bending, about 84 arcsec. The the best agreement with the expctations is found when the central region (2) is irradiated. These results can be explained in terms of depth of the stack traveled by the diffracted beam.

%
% Figure 11
%

\begin{table}[!h]
\caption{\footnotesize FWHM of the RCs performed} 
\label{tab:table3}
\centering
\medskip
\begin{tabular}{|c|c|c|} 
\hline
Region      & Distance from the  &  FWHM\\    
 analyzed   & grooved surface (mm) & (arcsec) \\    
\hline
 1 & 0.4 & 45.87 \\
\hline
 2 & 1.5 & 87.46 \\
 \hline
 3 & 2.3 & 79.12\\
 \hline

\end{tabular}
\end{table}

\textbf{Geometry 2 test}\\
%\label{sec:geometry2}

The stack in geometry 2 has been also tested. In this case the beam is incident on the grooved surface. In this case the diffracting planes are the (220), which 
do not show the $quasi$-$mosaic$ structure of the (111) planes. The thickness in this case is only 3 mm. Five regions of the cross-section have been irradiated.

%
% Figure 12
%
\begin{figure}[!h]
\begin{center}
\includegraphics[scale=0.3]{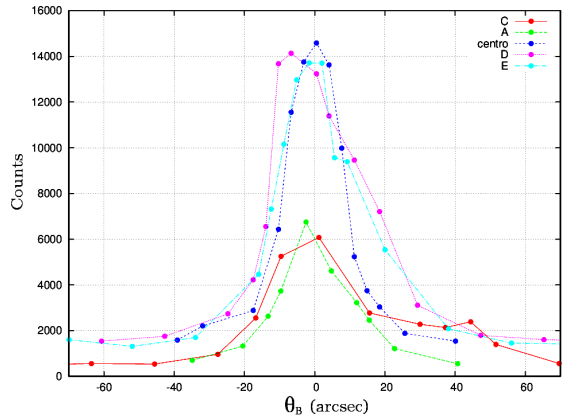}
\caption{\footnotesize  Diffracted counts (Geometry 2) from different regions (Fig.\ref{fe_stack}) of the stack.}
\label{5points}
\end{center}
\end{figure}

The test results are shown in Fig.~\ref{5points}. that layers of the stack are not perfectly aligned with each other, with the maximum misalignment being more relevant in the external regions (C and A).  

\subsubsection{Quasi-mosaic Silicon (111) samples in geometry 2}

\begin{figure}[!h]
\begin{center}
\includegraphics[scale=0.2]{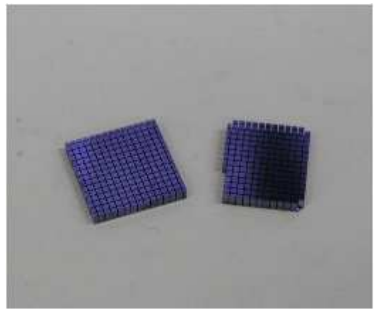}
\caption{\footnotesize  The grooved Silicon crystals tested in quasi-mosaic configuration.}
\label{f:quasi}
\end{center}
\end{figure}

We have characterized two samples (n5 and n6) with a square surface of $20\times 20$~mm$^2$ and thickness 
of 2 mm, with two different external curvature radii $R_e$: 60 and 8 meters, respectively (Fig. \ref{f:quasi}). 
The diffracting planes (111), in transmission configuration, are orthogonal to the square surface. From the dynamical 
theory of diffraction of bent crystals the ratio between the internal curvature radius R$_i$ and R$_e$ is equal to 2.6, with a consequent internal curvature of the diffracting 
(111) planes of 160 and 20 meters, respectively.

\begin{figure}[!h]
\begin{center}
\includegraphics[scale=0.25]{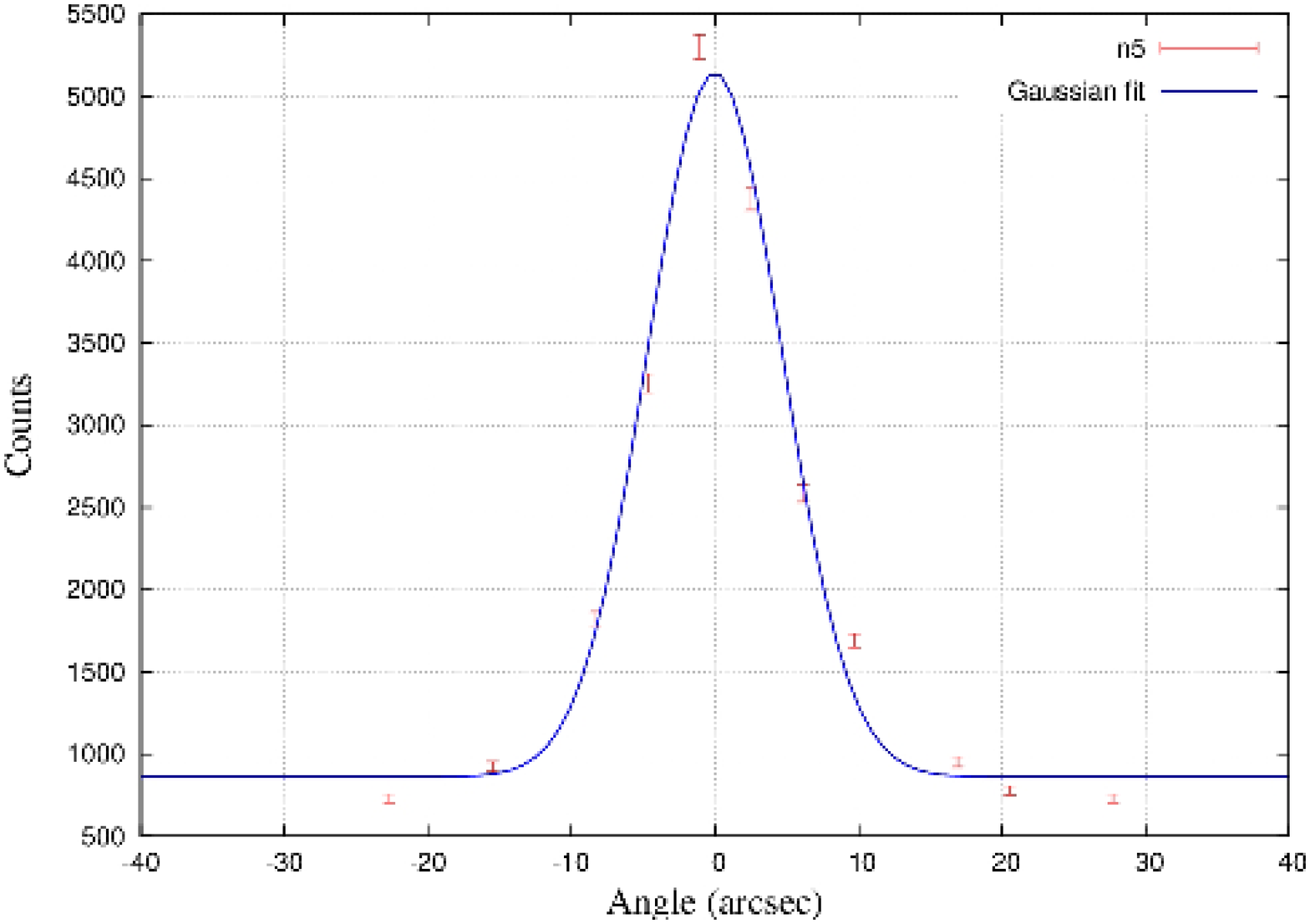}
\includegraphics[scale=0.25]{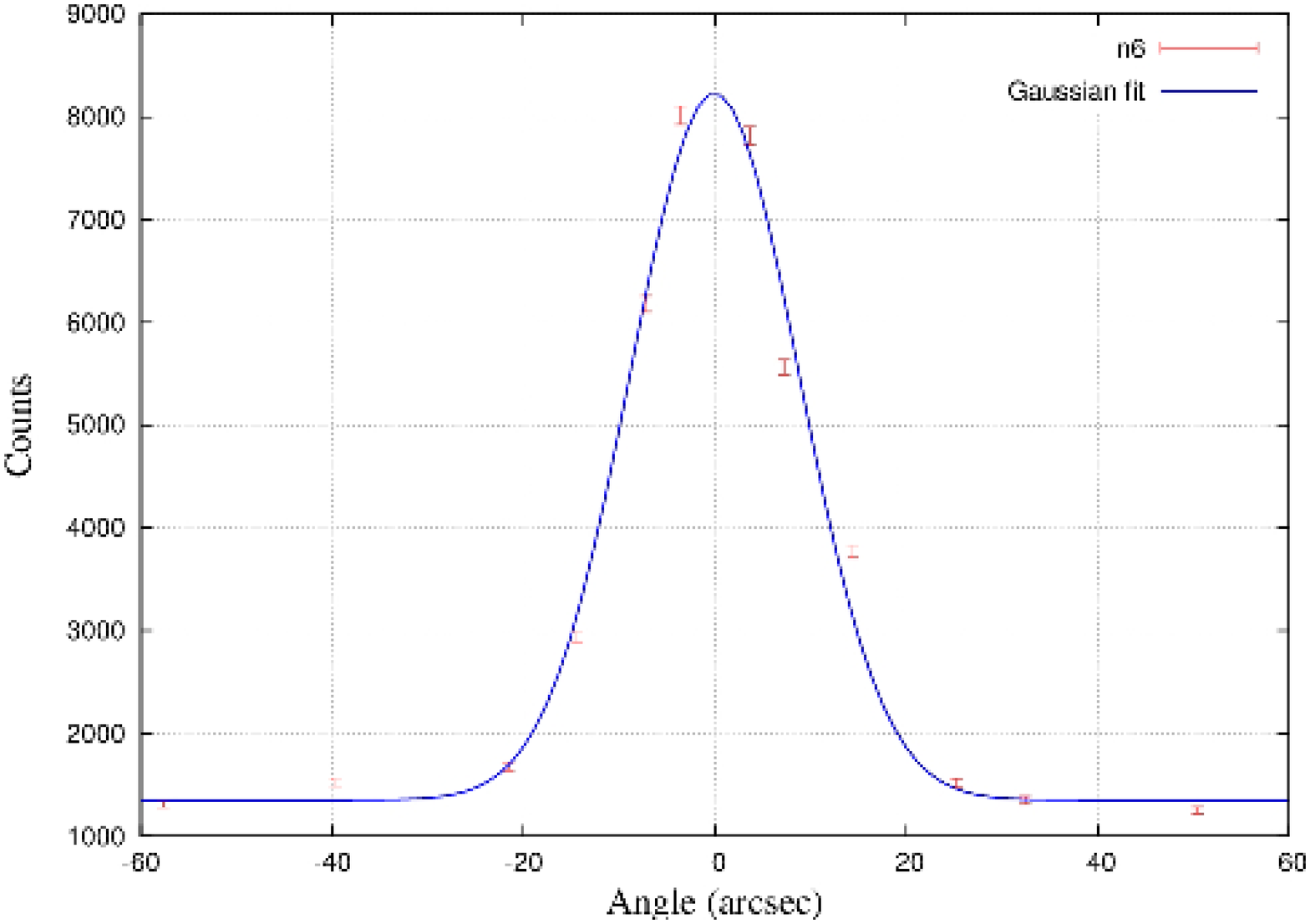}
\caption{\footnotesize  RCs of the Silicon samples n5 and n6 respectively.}
\label{n5n6}
\end{center}
\end{figure}

From the measured rocking curve we derived an angular spread of 11 $\pm$ 2 arcsec for sample n5. This value 
is in agreement with the convolution of the expected intrinsic RC width of the sample (2.5 arcsec) with the 
beam divergence. The measured reflectivity at 59.2 keV is about 80\%.  For the sample n6 we measured a RC width  of 21 $\pm$ 2 arcsec and a reflectivity value of 27\% (Fig.~\ref{n5n6}).

%
% Figure 13
%

%
% Figure14
%

\section{CONCLUSIONS}

In this paper, various bent and flat crystals have been tested for their possible use in Laue lenses.
For each sample, angular spread and diffraction efficiency at 59.2 keV have been measured as a function of the crystal thickness and curvature radius and compared with those obtained using flat crystals. 
We have also shown that the LARIX facility is suitable to perform these tests.

All the analyzed GaAs samples provided by IMEM show a
high-efficiency diffraction, with an average angular spread that satisfies
the requirements of the LAUE project. The curvature
radii are in agreement with the theoretical expectations.

The test of Silicon crystals provided by LSS with different curvature radii and
thicknesses have allowed us to establish that Silicon crystals could be successfully employed for LAUE project. However current technological limitations to bend crystals with thickness greater than 2 mm prevent to use  them.  Hence, Germanium tiles are more approriate to be used for the LAUE project, given that 2 mm thickness of this material is sufficent to achieve the needed diffration efficiency. Similarly, bent crystals of GaAs(111) in mosaic configurationcan be also used for the LAUE project.

\acknowledgments     
The crystals testing and the lens development are the result of a big effort of many people and institutions. 
We wish to thank in particular people of CNR/IMEM Institute in Parma, Sensor and of Semiconductor Laboratory (LSS) 
of the University of Ferrara for their close support, DTM in Modena, and INAF/IASF in Bologna. This work has been supported 
by the Agenzia Spaziale Italiana (ASI) through the project ''LAUE - $Una$ $Lente$ $per$ $i$ $raggi$ $Gamma$" under contract under contract I/068/09/0.

%%%%%%%%%%%%%%%%%%%%%%%%%%%%%%%%%%%%%%%%%%%%%%%%%%%%%%%%%%%%%
%%%%% References %%%%%

\bibliography{charact}   %>>>> bibliography data in report.bib
\bibliographystyle{spiebib}   %>>>> makes bibtex use spiebib.bst

\end{document}